\documentclass[12pt]{article}
\pdfoutput=1
\usepackage{geometry,enumerate,amsmath,amssymb,amsfonts}
\usepackage{fullpage}
\usepackage{graphicx}
\usepackage{caption}
\usepackage{subcaption}
\numberwithin{equation}{section}
\newcommand{\be}{\begin{equation}}
\newcommand{\ee}{\end{equation}}
\newcommand{\bea}{\begin{eqnarray}}
\newcommand{\eea}{\end{eqnarray}}
\newcommand{\bphi}{\mbox{\boldmath $\phi$}}

\renewcommand{\epsilon}{\varepsilon}

\begin{document}
\title{
\begin{flushright}\ \vskip -2cm {\small{\em DCPT-15/63}}\end{flushright}
\vskip 2cm 
The Dynamics of Aloof Baby Skyrmions}
\author{\ \\ Petja Salmi and Paul Sutcliffe\\[10pt]
{\em \normalsize Department of Mathematical Sciences,
Durham University, Durham DH1 3LE, U.K.}\\[10pt]
{\normalsize 
petja.salmi@gmail.com \quad\&\quad\  p.m.sutcliffe@durham.ac.uk}
}
\date{November 2015}
\maketitle
\begin{abstract}
  The aloof baby Skyrme model is a (2+1)-dimensional theory with
   solitons that are lightly bound. It is a low-dimensional analogue
   of a similar Skyrme model in (3+1)-dimensions, where the lightly bound solitons have binding energies comparable to nuclei.
   A previous study of static solitons in the aloof baby Skyrme model
   revealed that multi-soliton bound states have a cluster structure,
   with constituents that preserve their individual identities due to the
   short-range repulsion and long-range attraction between solitons.
   Furthermore, there are many different local energy minima that are all well-described by a simple binary species particle model.
   In this paper we present the first results on soliton dynamics in
   the aloof baby Skyrme model. Numerical field theory simulations
   reveal that the lightly bound cluster structure results in a variety of exotic soliton scattering events that are 
   novel in comparison to standard Skyrmion scattering.
   A dynamical version of the binary species point particle model is shown to provide a good qualitative description of the dynamics.
\end{abstract}

\newpage 
\section{Introduction}\quad
The Skyrme model \cite{Sk} is an effective low energy nonlinear
theory of pions, in which baryons correspond to solitons.
Given that the model has only a small number of parameters, 
it has been reasonably successful in providing a qualitative
description of a range of phenomena in nuclear physics \cite{BR}.
However, the standard Skyrme model fails to provide a good quantitative
match to some aspects of nuclei, particularly regarding
nuclear binding energies. A long-standing problem is that
Skyrmions are far too tightly bound \cite{BS3} in comparison to experimental data.
Recently, there has been some significant progress, with the
discovery of a variety of modified Skyrme models that
can alleviate the large binding energy problem. 
One promising example is the lightly bound model studied in \cite{GHS},
where numerical results show that it is indeed possible to select the
parameters of the (3+1)-dimensional theory in such a way 
that soliton binding energies are in reasonable quantitative agreement
with the experimentally known values for nuclei. 

The aloof baby Skyrme model, introduced in \cite{SS}, is the (2+1)-dimensional
analogue of the lightly bound Skyrme model. In both models,
the force between solitons exhibits repulsion at short range but attraction at 
long range. This creates a new length scale, associated with the optimum
distance
between solitons in the static multi-soliton cluster structures.
This new length scale can be tuned independently of the size of
a single soliton and determines the magnitude of soliton binding energies.
Static energy minimizing multi-soliton solutions have been computed in the aloof baby Skyrme model for soliton numbers from two to twelve \cite{SS}, together with a large number of stable local energy minima.
These additional static solutions often have energies that are very close to
those of the minimal energy solitons, indicating an intricate energy landscape.
A binary species point particle model, based on the assumption that soliton orientations are either maximally attractive or repulsive, has proved to be a useful approximation for studying these static solitons and their energy landscape. 

The purpose of the present paper is to provide the first results on soliton dynamics in the aloof baby Skyrme model. The main motivation is to provide an understanding of the type of dynamical phenomena that appear in lightly bound Skyrme models. In addition to performing numerical simulations of the full nonlinear field theory, we also introduce a dynamical version  
of the binary species point particle model and demonstrate that it
provides a reasonable approximate description of the dynamics.
Both aspects should be helpful for any future investigations of soliton dynamics in the related (3+1)-dimensional lightly bound Skyrme model \cite{GHS}, where studies to date have been limited to static solitons.

The paper is organised as follows. In section 2 we describe the model and briefly review some of the main results concerning static solitons and their point particle approximation using a binary species model. For a more detailed description of the aloof baby Skyrme model and its static solutions we refer the reader to \cite{SS}.
In section 3 we study the dynamics of two solitons and introduce our dynamical version of the point particle model. In section 4 we present results on the scattering of a single soliton on a higher charge soliton cluster and make a comparison with the predictions of the dynamical point particle model. Finally, in section 5 we present our conclusions.   

\section{The aloof baby Skyrme model}\quad
The aloof baby Skyrme model, introduced in \cite{SS}, is defined by the 
Lagrangian density
\begin{eqnarray}
\mathcal{L} \, =  \,
\frac{1}{2} \, \partial_{\mu} \boldsymbol{\phi} \cdot 
\partial^{\mu} \boldsymbol{\phi} 
-\frac{1}{4} \left( \partial_{\mu} \boldsymbol{\phi} \times 
\partial_{\nu} \boldsymbol{\phi} \right) \cdot
\left( \partial^{\mu} \boldsymbol{\phi} \times 
\partial^{\nu} \boldsymbol{\phi} \right) - {m^2}(1-\phi_3)(1+(1-\phi_3)^3) \, ,
\label{lagrangian}
\end{eqnarray}
where the field $\boldsymbol{\phi}=(\phi_1,\phi_2,\phi_3)$
is a three-component unit vector. The spacetime coordinates are $x_\mu$, with
$\mu=0,1,2$; and $m$ is the  
mass of the fields $\phi_1$ and $\phi_2,$ associated with
elementary excitations around the vacuum $\bphi=(0,0,1).$
To agree with the choice in \cite{SS}, we fix $m^2=0.05$ from now on.

Finite energy imposes the boundary condition that $\bphi$ tends to its vacuum
value at spatial infinity. This compactification implies that there is a
conserved 
integer-valued topological charge, or soliton number, given explicitly by
\be
N=-\frac{1}{4\pi}\int \bphi\cdot(\partial_1\bphi\times\partial_2\bphi)\, d^2x.
\label{charge}
\ee
This soliton number is equal to the degree of the map $\bphi$  from the compactified spatial plane to the target two-sphere.

The $N=1$ soliton, positioned at the origin, takes the radially symmetric form
\be
\bphi=(\sin f\, \cos(\theta+\chi),\,\sin f\, \sin(\theta+\chi),\, \cos f),
\label{radial}
\ee
where $\rho$ and $\theta$ are polar coordinates in the plane,
with $f(\rho)$ a monotonically decreasing radial profile function
satisfying the boundary conditions $f(0)=\pi$ and $f(\infty)=0.$ The arbitrary constant
angle $\chi$ is an internal phase associated with the global
$SO(2)$ symmetry of the Lagrangian density (\ref{lagrangian}),
that rotates the $\phi_1$ and $\phi_2$ components.
The energy of the 1-soliton is $E_1=20.27,$ which is obtained by numerically solving for the profile function $f(\rho).$

\begin{figure}[ht]
  \begin{center}
  \includegraphics[width=7cm]{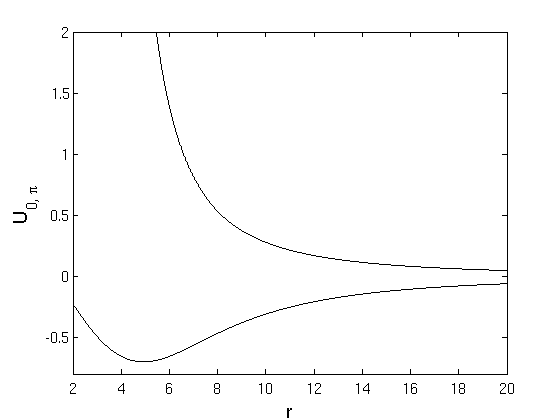}
  \includegraphics[width=7cm]{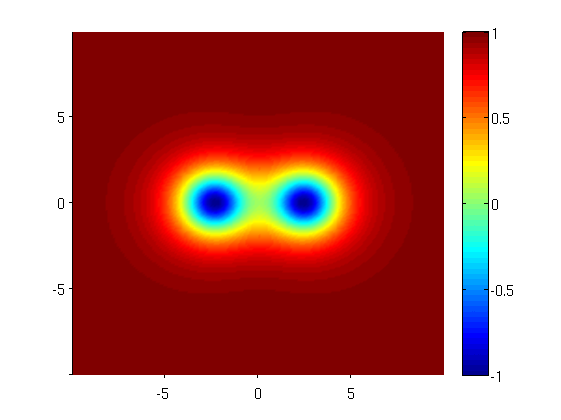}
    \caption{Left image: The interaction potentials $U_0$ (upper curve) and $U_\pi$ (lower curve). Right image: A plot of the field $\phi_3$ for the static 2-soliton. The size of the displayed area is $20\times 20$ and the colour coding shown here will be used throughout the paper.}
\label{fig-interaction}
  \end{center}
\end{figure}

The force between two well-separated solitons depends crucially on their
relative internal phase $\chi=\chi_1-\chi_2$, where $\chi_1$ and
$\chi_2$ are the phases of the two individual solitons. Let $E_\chi(r)$ denote
the energy of two $N=1$ solitons separated by a distance $r$ with relative phase
$\chi$. The interation energy is the difference between this energy and twice the energy of a single soliton, that is, it is given by $U_\chi(r)=E_\chi(r)-2E_1.$
In our earlier work \cite{SS} the interaction energy was computed numerically for $\chi=0$ (the repulsive channel) and $\chi=\pi$ (the attractive channel) and the data used to fit a Pad\'e approximant of order [3/4] in each case. These  Pad\'e approximants are plotted in the left image in Figure \ref{fig-interaction}, and we refer the reader to \cite{SS} for the associated explicit constants in the two Pad\'e approximants.

From this plot we see that two solitons that are in phase ($\chi=0$) are repulsive at all separations, whereas two solitons that are exactly out of phase ($\chi=\pi$) have a long-range attraction and a short-range repulsion. This produces an optimal separation $r=r_\star\approx 5$ at which the energy is minimized and found to be $E_2=39.84<2E_1.$ This minimal energy 2-soliton is displayed in the right image in Figure \ref{fig-interaction}, by plotting the field $\phi_3.$ Although the two solitons form a bound state they keep their individual identities, remaining aloof. The solitons are only lightly bound, with a binding energy of the order of one percent of the total energy. The locations of the component solitons are defined to be the points where $\bphi=(0,0,-1)$, which is the point on the target two-sphere that is antipodal to the vacuum value. It is evident from this figure that the separation between the two component solitons is indeed $r_\star\approx 5.$

Minimal energy solitons for larger values of $N$ have a similar cluster structure, composed of single solitons, and were studied in detail in \cite{SS} for $N\le 12$. Examples of particular relevance to the present work are the minimal energy $N$-solitons with $N=5,6,8,9$, which respectively take the form of a linear chain, a hexagon, an octagon and a $3\times 3$ square. In each case, all constituent solitons are exactly out of phase with all nearest neighbours, so that all these pairs of interactions are in the attractive channel. This property extends not only to all the minimal energy solitons, but also to the large number of local energy minima that exist in this theory and have energies that are only slightly above
the minimal energy values.

As we have seen, the attractive channel for two solitons is when the relative
phase between the two solitons is equal to $\pi.$ This suggests that 
the constituent single solitons in an $N$-soliton solution can be allocated
into two groups, where all solitons in a given group have the same phase
and there is a relative phase of $\pi$ between two solitons in different groups.
An appropriate arrangement of the solitons will then allow a large number of 
attractive channel pairings, with repulsive pairings being suppressed by
optimal spatial positioning. 

The binary species point particle approximation, introduced
in \cite{SS}, makes use of the
above property by considering $N$ particles, with positions
${\bf x}^{(1)},\ldots,{\bf x}^{(N)}$, such that each particle is either
blue or red, as follows.
Blue particles represent single solitons with an internal phase $\chi=0$
and red particles model solitons with an internal phase $\chi=\pi.$
Two particles of different colours are therefore in the attractive
channel and their interaction potential is given by $U_\pi(r),$ where
$r$ is the distance between the two particles. Similarly, two particles
of the same colour are in the repulsive channel and their interaction
potential is $U_0(r)$.
The total interaction energy for the particle system 
is the sum over all pairs of interactions, that is
\be
{\cal I}=\sum_{i=2}^N\sum_{j=1}^{i-1}U_{\chi_{ij}}(|{\bf x}^{(i)}-{\bf x}^{(j)}|),
\label{int}
\ee
where $\chi_{ij}$ is $0$ or $\pi$ depending on whether the particles with positions 
${\bf x}^{(i)}$ and ${\bf x}^{(j)}$ have the same or different colours.
Minimizing the energy ${\cal I}$ yields configurations of point particles
that are in remarkable agreement with the soliton results, upon
setting the difference between the number of blue and red particles to be $N$ mod 2.

In the following section we discuss the dynamics of two single solitons and describe a dynamical extension of the binary species point particle model.

\section{The dynamics of a soliton pair}\quad
To study soliton dynamics in the aloof baby Skyrme model we numerically solve the second order nonlinear field equations that follow from the variation of the  Lagrangian density (\ref{lagrangian}). A numerical approach to simulate soliton dynamics in the usual (3+1)-dimensional Skyrme model is discussed extensively in \cite{BS3} and we follow that method relatively closely here. Specifically, we use fourth-order accurate finite difference approximations for spatial derivatives
on a square grid containing $800^2$ lattice points, with a lattice spacing $\Delta x=0.1$ and time step $\Delta t=0.01.$ At the boundary of the grid the fields are set to the vacuum value $\bphi=(0,0,1).$ To reduce the radiation reflecting from the boundary of the grid we apply the adiabatic damping technique described in \cite{GS}, where damping on the field momenta  is applied at the lattice sites near the boundary of the grid. Note that in the figures that follow, we include only some relevant portion of the complete computational grid, which has an area of $80\times 80.$
 
\begin{figure}[ht]
\begin{center}
\includegraphics[width=3.4cm, height=3.2cm]{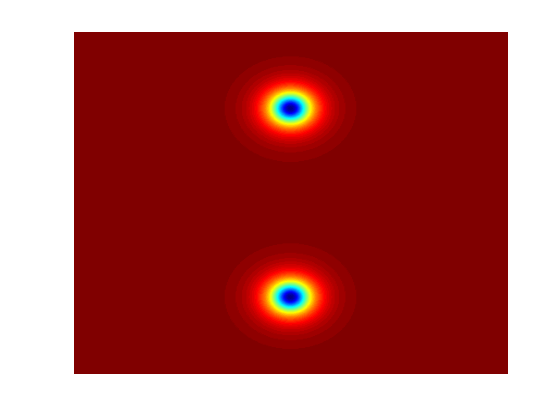}%
\includegraphics[width=3.4cm, height=3.2cm]{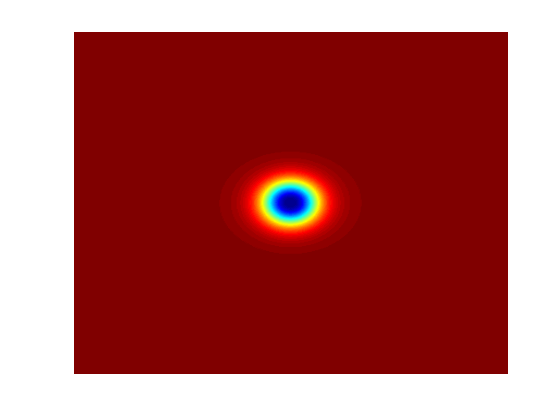}%
\includegraphics[width=3.4cm, height=3.2cm]{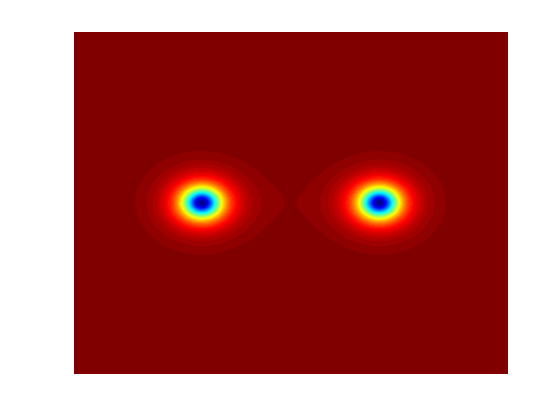}%
\includegraphics[width=3.4cm, height=3.2cm]{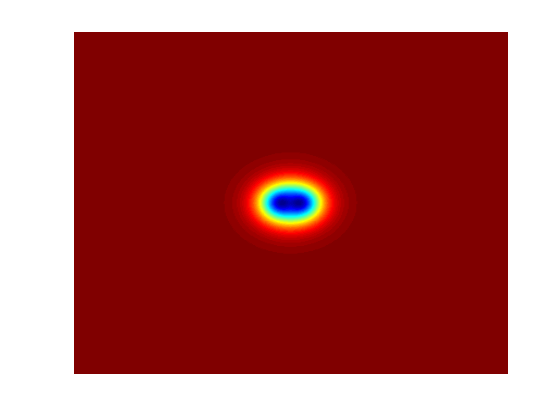}%
\includegraphics[width=3.4cm, height=3.2cm]{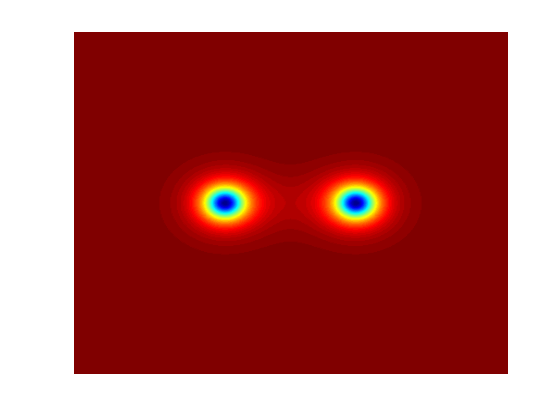}
   \caption{The head-on scattering of two solitons at $t=0,118,236,354,420$. 
     Each soliton has an initial speed $0.05$ and moves parallel to the $y$-axis. The size of the displayed area is $40\times 40.$}
   \label{fig-scat2}
\end{center}
\end{figure}

Two coincident solitons form a static radially symmetric solution that is unstable and has an energy that is very close to twice that of a single soliton. It is therefore to be expected that two well-separated single solitons in the attractive channel will coincide in a head-on collision, if given a small initial velocity towards each other. This is indeed the case, as demonstrated in Figure \ref{fig-scat2}, where at time $t=0$ (first image) each soliton has an initial speed $0.05$ and moves parallel to the $y$-axis. At $t=118$ (second image) the solitons are coincident and form an almost radially symmetric configuration. The two solitons then scatter at right angles to the initial direction of approach, as is familiar from the scattering of two solitons in the standard baby Skyrme model \cite{PSZ2}. As this interaction generates radiation, the two solitons no longer have enough energy to escape to infinity, with the given initial speed. They remain trapped, with an energy that is insufficient to allow another formation of coincident solitons, so no further right-angle scatterings can take place. The soliton pair now oscillate around the minimal energy separation (last three images), moving along a line perpendicular to the original direction of motion, until they eventually settle at the static 2-soliton solution shown in the right image in Figure \ref{fig-interaction}.

The right-angle scattering of two topological solitons is generic, and in these types of theories it is a consequence of the geometry of a finite-dimensional 
space of 2-soliton field configurations \cite{Ma,Wa}. Right-angle scattering is a solitonic phenomenon and is not captured by simple point particle dynamics. However, it turns out (see the following section) that such solitonic features don't play a significant role in aloof soliton scattering that involves higher charge soliton clusters. The reason is that a soliton cluster involves multiple solitons with phases that differ by $\pi$ between nearest neighbours. This means that a single soliton scattering on a cluster encounters both attractive and repulsive channel interactions, unlike the case of a pair of solitons, where all interations can be made attractive. The upshot is that any attractive interaction is screened by neighbouring solitons in the repulsive channel and the formation of  coincident solitons, that would result in right-angle scattering, appears to be non-generic, even in head-on scattering processes. Fortunately, this allows an approximate point particle treatment of the dynamics of aloof solitons.

\begin{figure}[ht]
\begin{center}
\includegraphics[width=9.0cm]{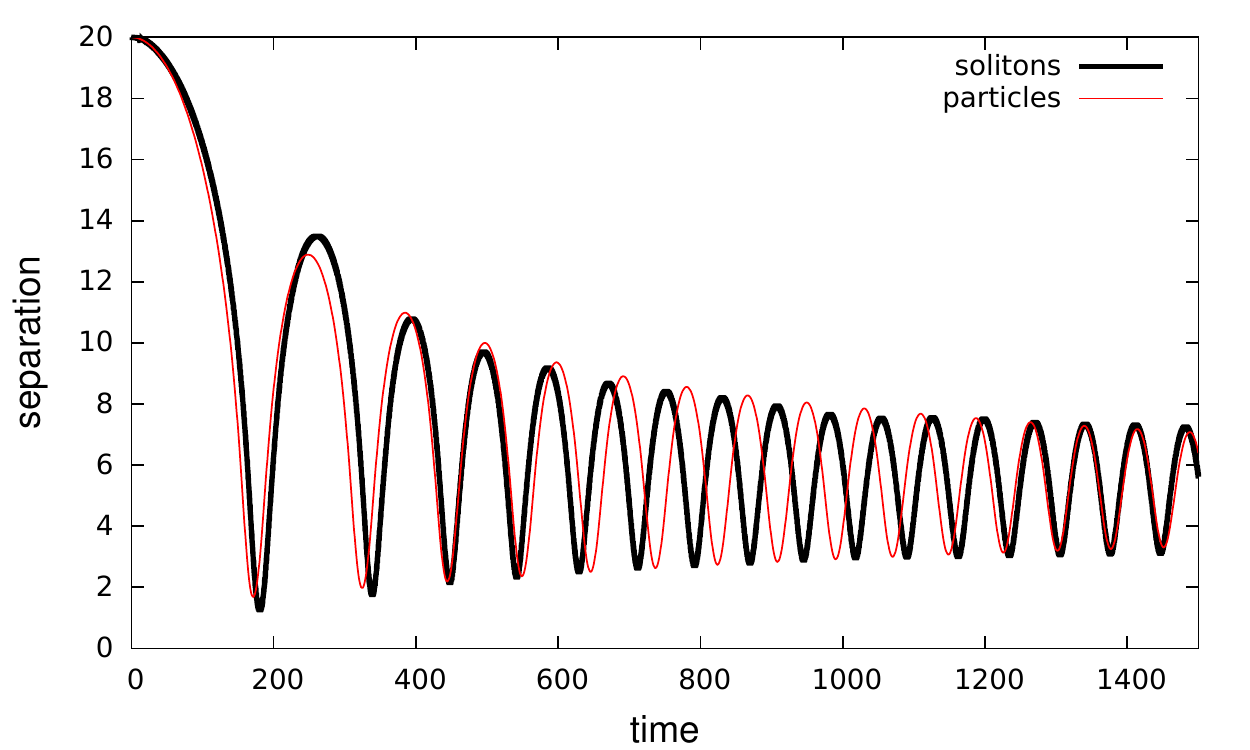}%
   \caption{The separation as a function of time for two initially static solitons in the attractive channel (thick black curve). The point particle approximation to this motion is also shown (thin red curve). 
}
   \label{fig-osc}
\end{center}
\end{figure}

Aloof solitons are characterised by small binding energies and a large number of local energy minima, that often have energies that are very close to those of the minimal energy solitons. This means that even relatively low energy scattering events may possess enough energy to explore a large part of the intricate energy landscape. Such explorations indeed take place and result in some exotic scattering processes, as described in section \ref{sec-cluster}. A crucial factor in determining the range of exploration is the amount of energy radiated by the solitons during their interactions with each other. To investigate the elasticity of soliton interactions we perform a simulation in which a pair of solitons initially at rest in the attractive channel are placed at a separation much larger than the optimal value $r_\star\approx 5.$ Figure \ref{fig-osc} displays the resulting evolution of the separation as a function of time (thick black curve) for an initial separation of $20.$ This figure shows that there is a reasonable decrease in the amplitude of the oscillation over the first cycle, when the separation gets significantly below the optimal value $r_\star$, but that later oscillations are fairly elastic, with only a very small decrease in the amplitude of the oscillation.

For any point particle model to reproduce the features of aloof soliton dynamics, even qualitatively, it must include a reasonable description of the energy lost by a soliton to radiation during interactions. The simplest way to incorporate such a mechanism is to include a damping term in the particle equations of motion, that depends upon the separation between the particles. As we shall only consider scattering events with modest soliton speeds, we ignore relativistic corrections and consider the following Galilean invariant equations of motion for the binary species point particles
\be
M\ddot {\bf x}^{(i)}+\kappa\sum_{j\ne i}e^{-\nu |{\bf x}^{(i)}-{\bf x}^{(j)}|}
(\dot {\bf x}^{(i)}-\dot{\bf x}^{(j)})=-\frac{\partial {\cal I}}{\partial {\bf x}^{(i)}}.
\label{particle}
\ee
The point particle mass is taken to be the energy of a static single soliton $M=E_1$ and ${\cal I}$ is the binary species interaction energy (\ref{int}). The constant $\kappa$ controls the strength of the damping term and we take a simple exponential dependence on the particle separation, with a scale parameter $\nu.$

We solve the system (\ref{particle}) numerically using a fourth-order Runge-Kutta method.
To fix the two parameters $\kappa$ and $\nu$ in the point particle model we compare the two particle case with the oscillating soliton pair displayed as the thick black curve in Figure \ref{fig-osc}. Fitting the results yields the parameters $\kappa=3.2$ and $\nu=1.25$, with the resulting particle approximation displayed as the thin red curve in Figure \ref{fig-osc}. This result shows that the simple particle model (\ref{particle}) indeed provides a reasonable description of both the oscillation period and the decay of the amplitude. The period is accurate to within around 10\%, but of course this means that over a large number of oscillations the particle approximation drifts in and out of phase with the soliton result. Nevertheless, this simple approximation is surprisingly accurate and should provide at least a qualitative description of soliton dynamics. In the following section we present some more complicated examples of soliton dynamics, involving scattering of solitons on higher charge clusters, and compare the results with the dynamical particle model. 

\section{Soliton scattering on a cluster}\quad\label{sec-cluster}
\begin{figure}[ht]
  \ \hskip 0.57cm
  \includegraphics[width=2.625cm]{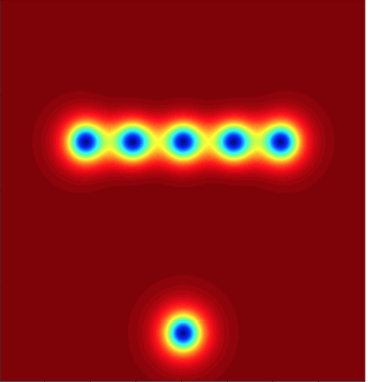} \hskip 1.26cm 
  \includegraphics[width=2.625cm]{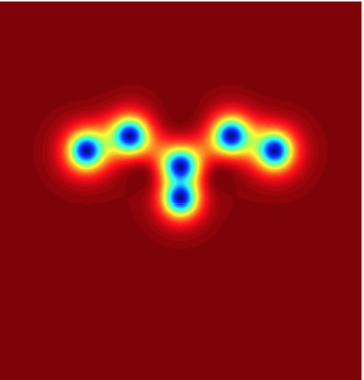} \hskip 1.26cm 
  \includegraphics[width=2.625cm]{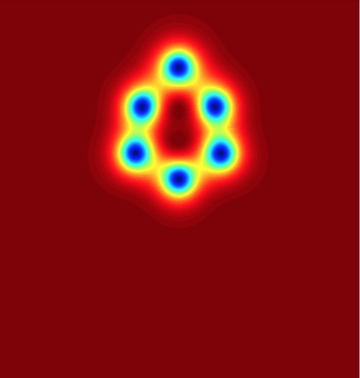} \hskip 1.26cm
  \includegraphics[width=2.625cm]{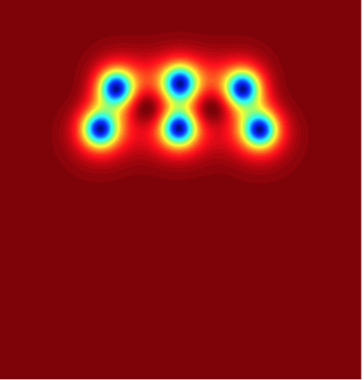} \hskip 1.26cm\\
  \includegraphics[width=3.9cm]{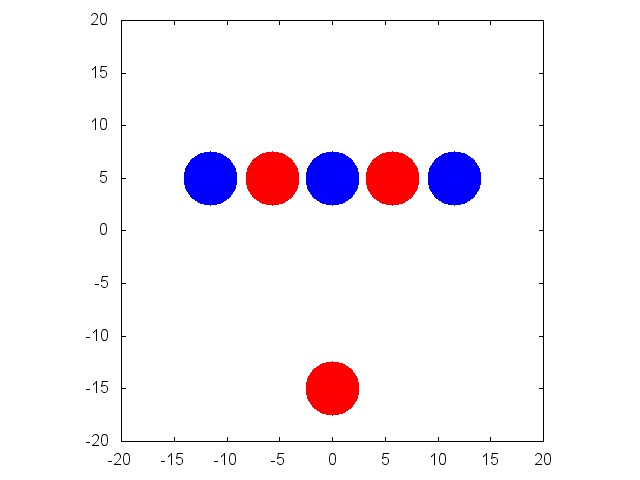}
  \includegraphics[width=3.9cm]{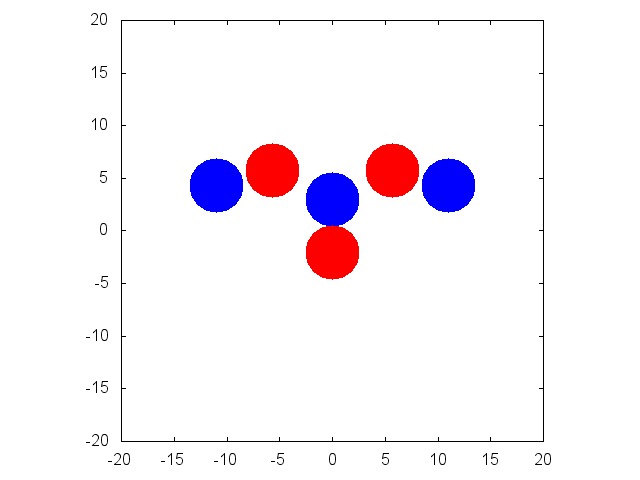}
  \includegraphics[width=3.9cm]{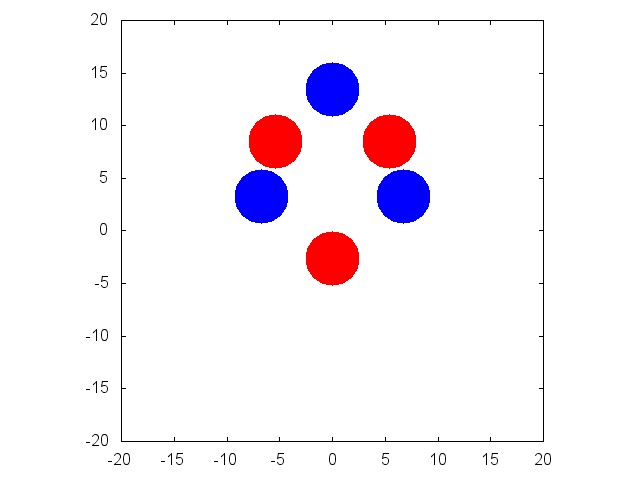}
  \includegraphics[width=3.9cm]{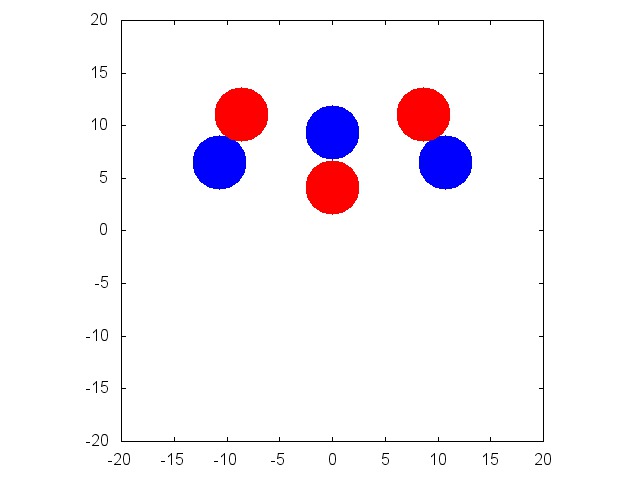}\\
   \caption{Solitons (top images) at $t=0,88,200,275$ and particles (bottom images) at $t=0,74,163,259.$
     The initial distance between the centre of mass of the 5-soliton and the 1-soliton is 20 and the 1-soliton has an initial speed 0.15.
     The size of the displayed area is $40\times 40.$}
   \label{fig-51}
\end{figure}

In this section we present some symmetric examples of the scattering of a single soliton on the initially static minimal energy $N$-soliton, with $N=5,8,9.$ We have performed a large number of simulations, but the chosen examples highlight the type of novel scattering phenomena that occur.

The minimal energy 5-soliton is a linear chain and can be seen in the first image in the top row of Figure \ref{fig-51}, together with a well-separated single soliton. The scattering event we investigate involves the single soliton moving with an initial speed 0.15 along a line perpendicular to the line of the chain. The 1-soliton is aimed at the middle of the chain in an attractive channel with the soliton that it is aimed at.
The first image in the bottom row of Figure \ref{fig-51} displays the corresponding initial point particles. The diameter of the discs is equal to the optimal separation $r_\star$ and the blue or red colour represents a phase of either 0 or $\pi.$
The phase of each soliton is not apparent in a plot of $\phi_3$ for the soliton image, but the point particle image makes it clear which soliton pairs are in the attractive channel (different colours) and which are in the repulsive channel (same colours). 

The additional images in Figure \ref{fig-51} show the resulting evolution of the solitons (top row) and the particles (bottom row). 
The solitons initially group into three pairs but subsequently form a hexagonal structure that is a deformation of the regular hexagon formed by the minimal energy 6-soliton. However, the hexagon then reverts to another triplet of pairs, with a more vertical alignment. This pattern of oscillation, between a hexagon and a triplet of pairs, continues with a large amplitude for much longer times than shown in Figure \ref{fig-51}, as one expects from the long-lived oscillations seen in Figure \ref{fig-osc} for a soliton pair. As this cluster of solitons is moving towards the boundary of the numerical grid then there is a time limit on the length of the field theory simulation, as boundary effects become significant as the solitons approach the edge of the grid. The amplitude of oscillation is still large at this time, so the field theory simulations cannot be followed long enough to reveal a rigid moving cluster. However, given the oscillation is around the minimal energy hexagonal 6-soliton, then the expectation is that the final state of this scattering is a hexagonal 6-soliton moving with constant speed.

There is no boundary limitation in the simulations of the dynamics of the point particle model and an evolution to $t=20000$ indeed reveals a hexagonal 6-soliton as the final outcome of this scattering event. As demonstrated by the bottom row of images in Figure \ref{fig-51}, the particle model provides a good qualitative description of the soliton dynamics, but the exact time scale is not reproduced: note that the snapshots are shown at different times for the soliton and particle systems. It is not surprising that the particle model has some quantitative deficiencies, given its simplistic form and inclusion of only 2-particle interactions and fixed relative phases, but it does provides a useful qualitative approximation. In particular, it correctly predicts the different conformations obtained in the scattering process and the order in which they appear.

The fact that the scattering of a 1-soliton on the minimal energy 5-soliton ultimately produces the minimal energy 6-soliton is not a novel feature for soliton dynamics. However, the mechanism for this production is unusual, in that it involves a very long-lived and large amplitude oscillatory state. This is a characteristic feature of aloof solitons, due to their low binding energies and plethora of local energy minima, and clearly has implications for the formation of resonance states in nuclear scattering described by a (3+1)-dimensional aloof Skyrme model. 
\begin{figure}[ht]
  \ \hskip 0.57cm
  \includegraphics[width=2.625cm]{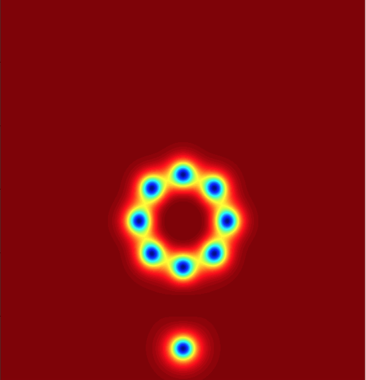} \hskip 1.26cm 
  \includegraphics[width=2.625cm]{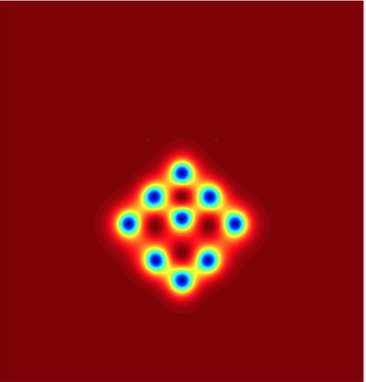} \hskip 1.26cm 
  \includegraphics[width=2.625cm]{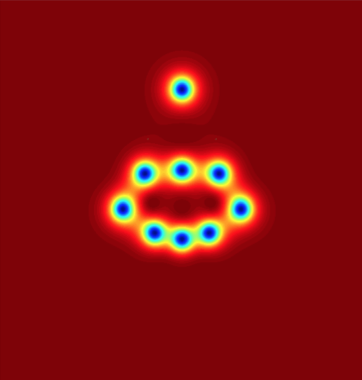} \hskip 1.26cm
  \includegraphics[width=2.625cm]{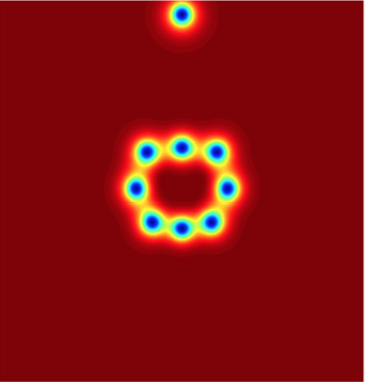} \hskip 1.26cm\\
  \includegraphics[width=3.9cm]{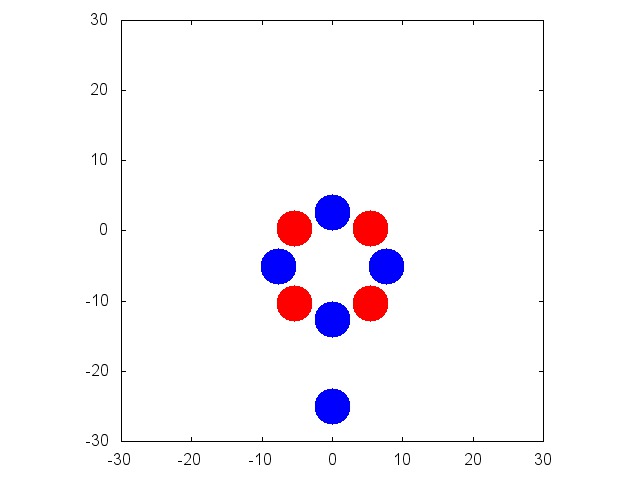}
  \includegraphics[width=3.9cm]{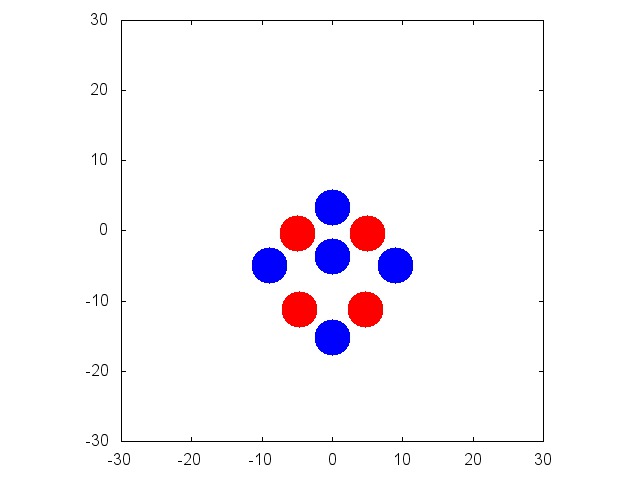}
  \includegraphics[width=3.9cm]{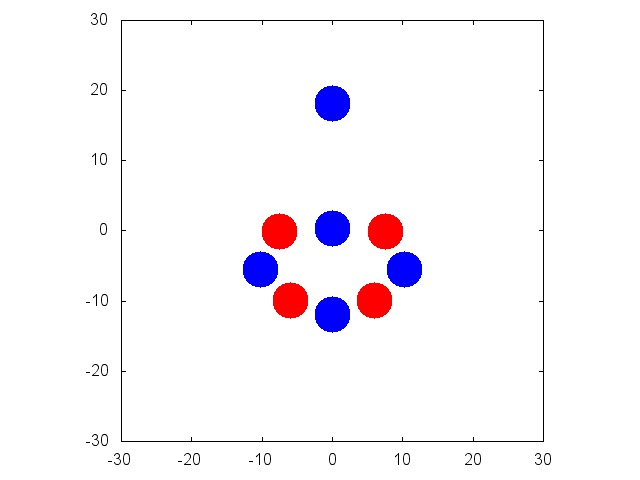}
  \includegraphics[width=3.9cm]{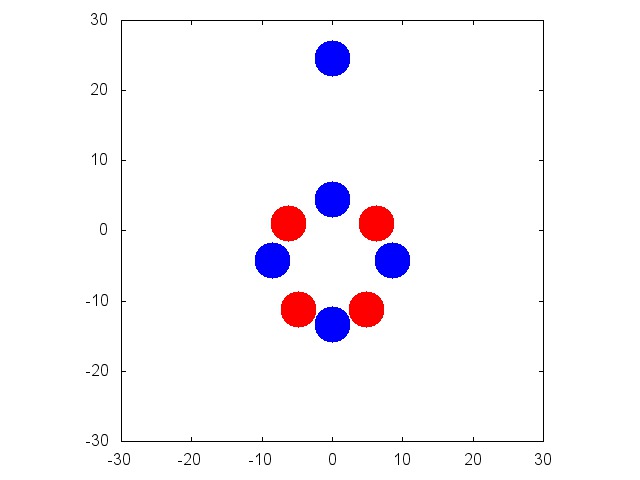}\\
  \caption{Solitons (top images)
    at $t=0,48,193,297$ and particles (bottom images) at $t=0,48,116,149.$
    The initial distance between the centre of mass of the 8-soliton and the 1-soliton is 20 and the 1-soliton has an initial speed 0.35.
    The size of the displayed area is $60\times 60.$}
     \label{fig-81}
\end{figure}

In Figure \ref{fig-81} we provide an illustrative example in which a single soliton scatters on a cluster in the repulsive channel. Again the top row of images are the soliton field theory computations and the botton row of images are the results from the particle approximation. As in the previous example, the two sets of images are  shown at different times to highlight the qualitative correspondence between the two systems, despite the fact that the timings disagree.
The selected example is the scattering of a single soliton on the minimal energy 8-soliton, which is a regular octagon of solitons with alternating phases that differ by $\pi.$ The 1-soliton has an initial speed of 0.35 and is in the repulsive channel with the soliton in the octagon that it is directed towards. This is clearly seen in the first point particle image, where both these solitons are blue, representing the same phase and hence repulsion between them.

\begin{figure}[ht]
  \ \hskip 0.57cm
  \includegraphics[width=2.625cm]{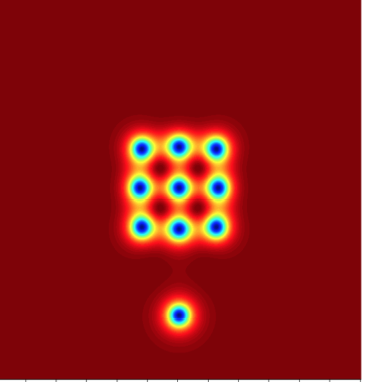} \hskip 1.26cm 
  \includegraphics[width=2.625cm]{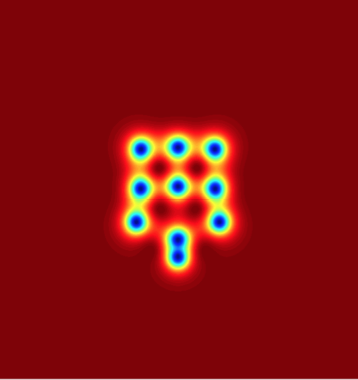} \hskip 1.26cm 
  \includegraphics[width=2.625cm]{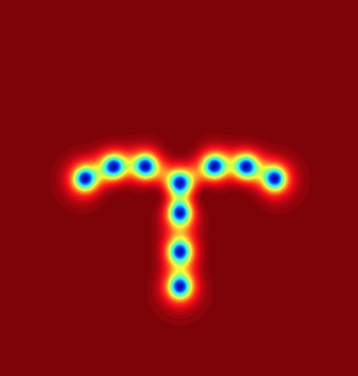} \hskip 1.26cm
  \includegraphics[width=2.625cm]{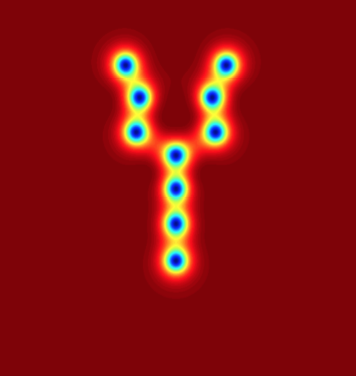} \hskip 1.26cm\\
  \includegraphics[width=3.9cm]{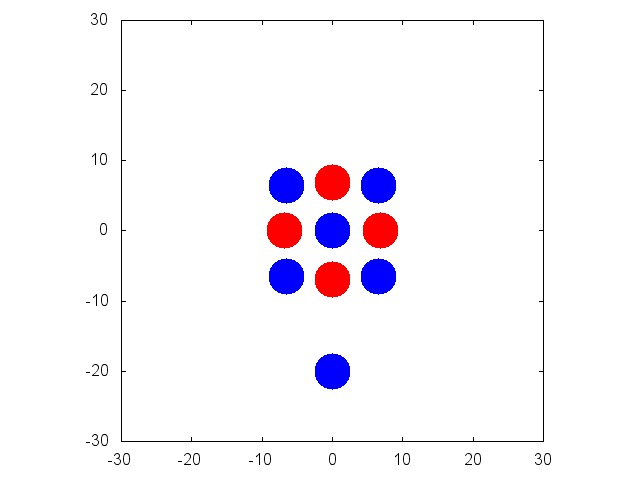}
  \includegraphics[width=3.9cm]{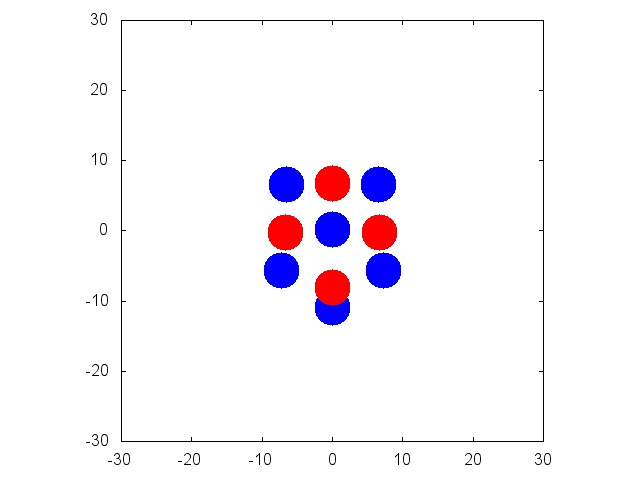}
  \includegraphics[width=3.9cm]{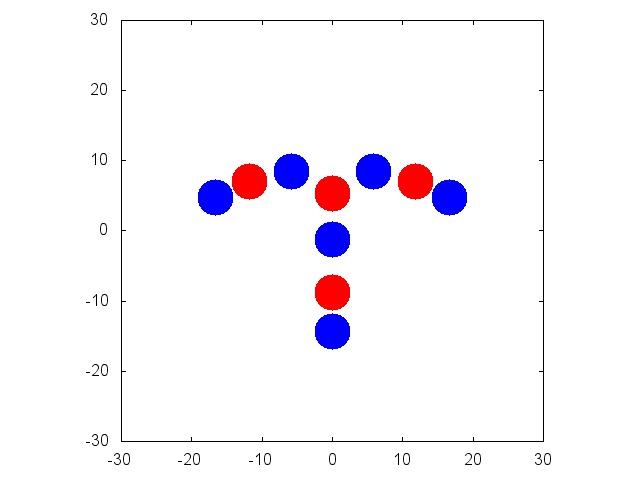}
  \includegraphics[width=3.9cm]{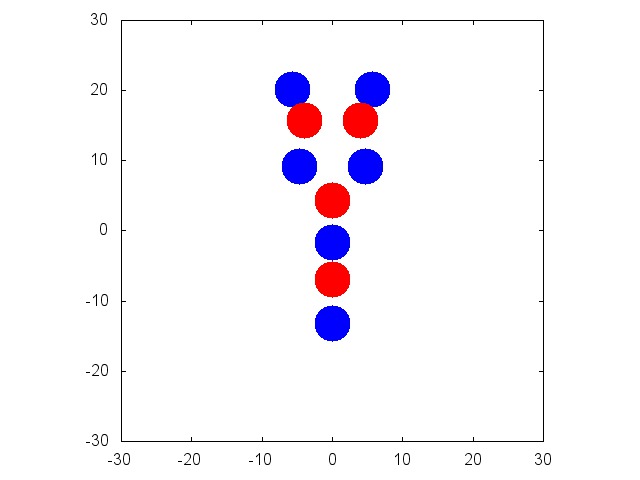}\\
  \caption{Solitons (top images)
    at $t=0,38,235,600$ and particles (bottom images) at $t=0,38,166,372.$
    The initial distance between the centre of mass of the 9-soliton and the 1-soliton is 20 and the 1-soliton has an initial speed 0.25.
  The size of the displayed area is $60\times 60.$}
   \label{fig-91}
\end{figure}

The resulting evolution displayed in Figure \ref{fig-81} is a novel scattering process in which the first stage is the formation of a configuration that is close to the minimal energy 9-soliton: a $3\times 3$ square of nine solitons with each nearest neighbour pairing in the attractive channel (see the first image in Figure \ref{fig-91}). This is created by the incoming soliton pushing the soliton it approaches into the centre of the octagon and taking its place in the outer ring. The momentum transferred to the inner soliton means that it continues through to the opposite side of the cluster where it pushes a soliton from the edge of the structure and takes its place. This results in the escape of a 1-soliton that leaves behind a deformed and oscillating octagon. The octagon eventually relaxes to the minimal energy 8-soliton, so in terms of initial and final states only, an observer may be fooled into believing that the single soliton has simply passed straight through the 8-soliton. However, as we can see from both the soliton simulations and the point particle approximation, the dynamics is novel and more interesting than the corresponding in and out states might suggest. In particular, the 1-soliton that escapes in the out state is not the same 1-soliton that appears in the in-state. In typical soliton scattering events it is usually not possible to make statements about the fate of an individual soliton because solitons merge and it is not well-defined to assign a position to any given soliton throughout the scattering process. However, as aloof solitons generally remain distinct throughout a scattering event then we are able to refer to the evolution of individual solitons.  

The next process we illustrate is the formation of an excited state of a static solution that is only a local energy minimum rather than the global minimal energy soliton. Figure \ref{fig-91} displays the attractive channel scattering of a 1-soliton with speed $0.25$ on the minimal energy 9-soliton (a $3\times 3$ square). This results in a highly excited state that is a large amplitude oscillation around a static charge 10 solution with triangular $C_3$ symmetry, in which three arms each containing three solitons with alternating phases are joined to a central soliton. The static $C_3$ symmetric charge 10 solution is stable, but it has a slightly higher energy than the regular decagon that is the $C_{10}$ symmetric 10-soliton with minimal energy. The large amplitude oscillation of the $C_3$ symmetric solution can be seen in the third and fourth images in Figure \ref{fig-91} (both top and bottom) and corresponds to a flapping of two of the three arms.    

\begin{figure}[ht]
\centering
  \includegraphics[width=3.625cm]{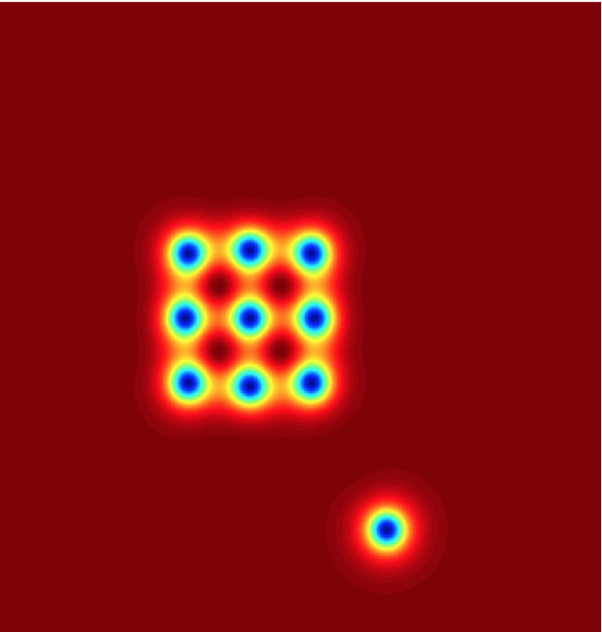} 
  \includegraphics[width=3.625cm]{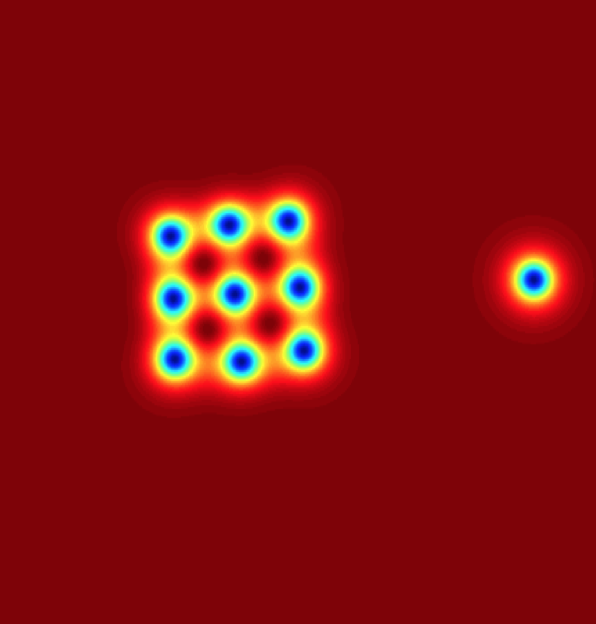} \hskip 0.16cm\\ 
  \includegraphics[width=3.625cm]{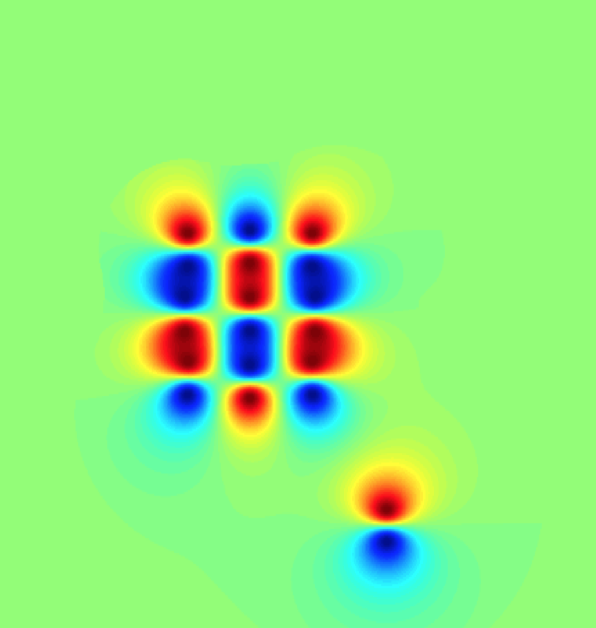} 
  \includegraphics[width=3.625cm]{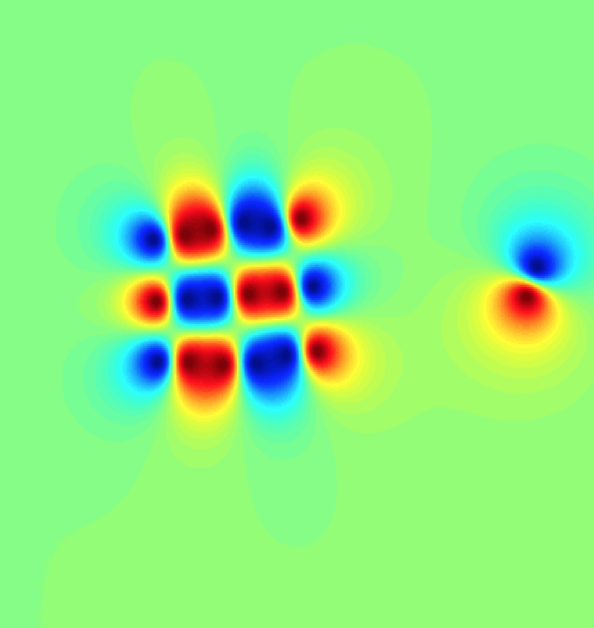} \hskip 0.16cm \\
   \caption{The field $\phi_3$ (top images) and $\phi_2$ (bottom images) at $t=0$ (left images) and $t=285$ (right images) when a single soliton with initial speed $0.15$ is deflected by the initially static 9-soliton. The size of the displayed area is $60\times 60.$}
\label{fig-spin}
\end{figure}

The examples presented in this section illustrate the novel scattering processes that arise for aloof solitons and suggest general features that are expected to appear in similar models in (3+1)-dimensions. A particularly distinctive feature is the formation of 
large amplitude and long-lived oscillatory states based on vibrational modes of stable static arrangements of solitons, including (but not limited to) minimal energy solitons. 
We have demonstrated that a simple dynamical version of a binary species point particle model can provide a good qualitative description of aloof soliton dynamics. It is rather surprising that such a simple model works so well, in particular given that it assumes that all relative phases between solitons are either zero or $\pi.$ The reason that this appears to be a sufficient approximation is that neighbouring solitons remain locked to the out of phase attractive channel even when perturbed.

 To illustate this behaviour, we consider the evolution presented in Figure \ref{fig-spin}, in which a single soliton with initial speed 0.15 moves parallel to the $y$-axis, but with a large impact parameter that means it is not on a collision course with the static mininal energy 9-soliton. The 1-soliton is in the repulsive channel with the bottom right soliton in the $3\times 3$ square and hence is deflected away from the 9-soliton. This is shown in the top two images by plotting the field $\phi_3.$
 To observe the phases of the solitons, the images in the bottom row display the field  $\phi_2$ at the corresponding times. In these plots, red denotes a positive value of $\phi_2$, blue a negative value and green is where $\phi_2$ is close to zero. The $\phi_2$ field of a radially symmetric single soliton has a dipole pattern, with a red and a blue region, and the internal phase is the angle between the blue to red axis and the $y$-axis. The attractive channel between two solitons, where the relative phase is $\pi$, corresponds to two solitons overlapping a region of the same colour in plots of $\phi_2$. In particular, the left image in the bottom row of Figure \ref{fig-spin} confirms that all neighbouring pairs in the static 9-soliton are in the attractive channel. In the second plot in the bottom row (after the 1-soliton has been deflected) the 9-soliton has been perturbed by the interaction and we observe that all the phases of the solitons within the perturbed 9-soliton have changed by approximately $\pi/2$. However, the relative phases between all neighbouring pairs remain equal to $\pi$, as the soliton pairs remain locked in the attractive channel. Thus, although internal phases can vary, which is not captured by the particle model, relative phases between solitons in close proximity remain equal to zero or $\pi$, so the simple particle model can capture this qualitative behaviour.
Note that the deflected 1-soliton does not have a relative phase of zero or $\pi$ with respect to any of the solitons in the 9-soliton cluster.
However, it is not a concern that the point particle model fails to capture this aspect, at least for the issues studied in this paper, because solitons
that are far from each other have a negligible interaction. If well-separated solitons eventually move into close proximity then they will again lock to the out of phase attractive channel, which is within the description of the particle model.

\section{Conclusion}\quad
In this paper we have studied the dynamics of aloof baby Skyrmions using both field theory simulations and a point particle approximation. The numerical field theory studies demonstrate that aloof solitons generally retain their individual identities and hence it is possible to follow individual solitons through the scattering process, which is not possible for standard solitons.
The dynamics involving the scattering of a single soliton on a cluster is rather exotic. The observed outcomes include the replacement of one soliton by another and the formation of larger oscillating clusters, which may or may not be perturbed versions of the global minimal energy soliton.  Furthermore, the point particle model correctly reproduces the qualitative features of all these phenomena
and is useful for predicting the final state of scattering events where long relaxation times make field theory simulations prohibitive.
We expect similar dynamical behaviour in the (3+1)-dimensional version of the theory \cite{GHS} and it would be interesting to investigate this via field theory simulations and to predict the results with a point particle approximation.

Finally, we have seen that the scattering of aloof solitons can generate internal phase rotations even when the solitons remain quite distinct. These internal phase rotations are the analogues of isospin rotations in the (3+1)-dimensional Skyrme model and therefore the aloof baby Skyrme model may be a good system in which to investigate the ideas on the spin-orbit force described recently in \cite{HM}.
\section*{Acknowledgements} \quad
Many thanks to Peter Bowcock, David Foster, Derek Harland and 
Paul Jennings for useful discussions. This work is funded by the EPSRC grant EP/K003453/1 and the STFC grant ST/J000426/1.

\end{document}